\documentclass[aps,pre,showpacs,showkeys,twocolumn,groupedaddress]{revtex4}

\usepackage{graphicx}

\begin{document}

\newcommand {\ab} {{\bf a}}
\newcommand {\abd} {{\bf a^\dag}}
\newcommand {\xb} {{\bf x}}
\newcommand {\xbd} {{\bf x^\dag}}
\newcommand {\yb} {{\bf y}}
\newcommand {\ybd} {{\bf y^\dag}}
\newcommand {\xn} {{\bf x}_n}
\newcommand {\xnd} {{\bf x}_n^{\bf\dag}}
\newcommand {\yn} {{\bf y}_n}
\newcommand {\ynd} {{\bf y}_n^{\bf\dag}}
\newcommand {\zb} {{\bf z}}
\newcommand {\zbd} {{\bf z}^{\bf\dag}}
\newcommand {\ree} {R_\epsilon\left(E_1,E_2\right)}
\newcommand {\Str} {{\rm Str}}
\newcommand {\tr} {{\rm tr}}
\newcommand {\Sdet} {{\rm Sdet}}
\newcommand {\diag} {{\rm diag}}

\newcommand {\tb}{t_{12}}
\newcommand {\tc}{t_{21}}

\title{Fidelity freeze for a random matrix model with off--diagonal perturbation}

\author{H.-J. St\"ockmann\dag\ and H. Kohler\ddag}
\email{stoeckmann@physik.uni-marburg.de}
\email{kohler@tphys.uni-heidelberg.de}
\affiliation{\dag Fachbereich Physik der Philipps-Universit\"at
Marburg, Renthof 5, D-35032 Marburg, Germany\\
\ddag Institut f\"ur theoretische Physik, Universit\"at
         Heidelberg, Philosophenweg 19, D-69120 Heidelberg, Germany}

\date{\today}

\begin{abstract}

The concept of fidelity has been introduced to characterize the
stability of a quantum-mechanical system against perturbations. The
fidelity amplitude is defined as the overlap integral of a wave
packet  with itself after the development forth and back under the
influence of two slightly different Hamiltonians. It was shown by
Prosen and \v{Z}nidari\v{c} in the linear-response approximation
that the decay of the fidelity is frozen if the Hamiltonian of the
perturbation contains off-diagonal elements only. In the present
work the results of Prosen and \v{Z}nidari\v{c} are extended by a
supersymmetry calculation to arbitrary strengths of the perturbation
for the case of  an unperturbed Hamiltonian taken from the
Gaussian orthogonal ensemble and a purely unitary antisymmetric
perturbation. It is found that for the exact calculation the freeze
of fidelity is only slightly reduced as compared to the
linear-response approximation. This may have important consequences
for the design of quantum computers.

\end{abstract}

\pacs{05.45.Mt, 03.65.Yz,03.67.Lx}
\keywords{Fidelity, Random matrix theory, Quantum Computation}

\maketitle

\section{Introduction}

The concept of fidelity was originally introduced by Peres
\cite{per84} to characterize the quantum-mechanical stability of a
system against perturbations. Recently it enjoys a renewed
popularity because of its obvious relevance for quantum computing.
In the present context the works which focus on random matrix
aspects are of particular relevance. First the paper by Gorin et
al. \cite{gor04} has to be mentioned where the Gaussian average of
the decay of the fidelity amplitude was calculated in linear
response approximation. For small perturbations the authors found
a predominantly Gaussian decay, with a cross-over to exponential
decay for strong perturbations, in accordance with literature
\cite{pro02b,cer02}. The results of the paper could be
experimentally verified in an ultrasound experiment \cite{lob03b}
and in a microwave billiard \cite{sch1}. Using supersymmetry
techniques, the limitations of the linear response approximation
could be overcome, yielding analytic expressions for the decay of
the fidelity amplitude for the Gaussian orthogonal (GOE) and
Gaussian unitary (GUE) ensemble. Quite surprisingly a recovery of
the fidelity was found at the Heisenberg time \cite{stoe04b} which
was interpreted as a spectral analogue of an Debye-Waller factor
\cite{stoe05}. Reference \onlinecite{stoe04b} is the basis for the
present work.

The Gaussian decay observed for small perturbation is caused by
the diagonal part of the perturbation in the eigenbasis of the
unperturbed Hamiltonian. This was the motivation for Prosen and
\v{Z}nidari\v{c} to look for perturbations with zero diagonal,
first in classically integrable systems \cite{pro03c}. Later on
they extended there studies to classically chaotic systems
\cite{pro05}. In linear response approximation they found an
plateau in the decay of the fidelity. Only after extraordinarily
long times the decay started again, exponentially below, and
Gaussian beyond the Heisenberg time. It remained, however, an open
question whether this freeze of the fidelity is reality or whether
it is just an artifact of the approximation. It was the motivation
for the present work to answer this question by extending the
previous supersymmetry calculation to the freeze situation. It
will be shown that the freeze is present also in the exact
calculation. This may have important consequences for quantum
computing. If one succeeds in imbedding the atoms representing the
qbits into an environment coupled only via an off-diagonal
perturbation to the atoms, an enhancement of the system's
stability by orders of magnitude is expected.

The present results are not restricted to random matrices. In
reference \cite{gor} it is shown that, e.\,g., kicked tops with a
corresponding dynamics follow exactly the random matrix predictions
of the present paper.

\section{The linear response approximation}

The fidelity amplitude is defined as the overlap integral of an
initial wave function $|\psi\rangle$ with itself after the time
evolution due to two slightly different Hamiltonians $H_0$ and
$H_\lambda=H_0+\lambda V$,

\begin{equation}\label{l01}
    f_{\lambda}(\tau)=\left<\psi\left|e^{2\pi\imath
  H_\lambda\tau}e^{-2\pi\imath H_{0}\tau}\right|\psi\right>\,,
\end{equation}
where the time $\tau$ is  given in units of the Heisenberg time.
For chaotic systems $f_{\lambda}(\tau)$ is independent on the
initial condition, and we may replace expression (\ref{l01}) by
its average over $|\psi\rangle$, i.\,e.,
\begin{equation}\label{l02}
  f_\lambda(\tau)=\frac{1}{N}\left<{\rm tr}\left[e^{2\pi\imath H_\lambda\tau}
  e^{-2\pi\imath H_0\tau}\right]\right>\,,
\end{equation}
where $N$ is the rank of the Hamiltonians, and the brackets denote
an ensemble average. Under the assumptions that (i) $H_0$ is taken
either from the Gaussian orthogonal (GOE) or the Gaussian unitary
ensemble (GUE) with a  mean level spacing of one in the band
centre and (ii) that the variances of the matrix elements of $V$
are given by
\begin{equation}\label{l03}
    \left<V_{ij}V_{kl}\right>=\left\{\begin{array}{lc}
    \delta_{ik}\delta_{jl}+\delta_{il}\delta_{jk}\,,\qquad & {\rm(GOE)} \\
    \delta_{il}\delta_{jk}\,, &
    {\rm(GUE)}
\end{array}\right.
\end{equation}
Gorin et al. \cite{gor04} obtained for the fidelity amplitude in
the linear response approximation
\begin{equation}\label{l04}
  f_{\epsilon}(\tau)\sim 1- \epsilon \, C(\tau)\,,
\end{equation}
where $\epsilon=4\pi^2\lambda^2$, and $C(\tau)$ is given by
\begin{equation}\label{l05}
    C(\tau)=\frac{\tau^2}{\beta}+\frac{\tau}{2}-\int_0^\tau \int_0^t
b_{2,\beta}(t^\prime) dt^\prime dt \,.
\end{equation}
$b_{2,\beta}(\tau)$ is the two-point form factor, and $\beta$ is
the universality index, i.\,e. $\beta=1$ for the GOE, and
$\beta=2$ for the GUE. For the Gaussian ensembles
$b_{2,\beta}(\tau)$ is known, and $C(\tau)$ can be explicitely
calculated \cite{gor04}. The range of validity of the linear
response approximation can be somewhat extended, by exponentiating
Eq. (\ref{l04}),
\begin{equation}\label{l06}
  f_{\epsilon}(\tau)= e^{-\epsilon \, C(\tau)}\,.
\end{equation}
The authors argued that the errors of the approximation should be
fairly small for $\lambda\sim 0.1$ and negligible for$\lambda\sim
0.01$ (corresponding to $\epsilon=0.4$ and 0.004, respectively),
which was fully confirmed by the exact calculations
\cite{stoe04b}.

The Gaussian decay for small perturbations is caused by the
diagonal part of the perturbation. This is immediately evident
from Eq. (\ref{l02}): For small perturbations $V$ can be truncated
to $V_{\rm diag}$, its diagonal part in the basis of
eigenfunctions of $H_0$. In this regime Eq. (\ref{l02}) reduces to
%
\begin{eqnarray}\label{l07}
  f_\lambda(\tau)&=&\frac{1}{N}\left<{\rm tr}e^{2\pi\imath (H_0+\lambda V_{\rm diag})\tau}
  e^{-2\pi\imath H_0\tau}\right>\nonumber\\
  &=&\frac{1}{N}\left<\sum\limits_n e^{2\pi\imath (E_n+\lambda V_{nn})\tau}e^{-2\pi\imath E_n\tau}\right>
\nonumber\\&=&\frac{1}{N}\left<\sum\limits_n e^{2\pi\imath \lambda
V_{nn}\tau}\right>
  =\frac{1}{N}\left<{\rm tr}\ e^{2\pi\imath \lambda V_{\rm diag}\tau}\right>
  \nonumber\\&=&e^{-\frac{\epsilon}{2}\tau^2\left<V_{\rm diag}^2\right>}\,,
\end{eqnarray}
where the $E_n$ are the eigenenergies of $H_0$ \cite{cer02}. This
suggests to consider perturbations with vanishing diagonal matrix
elements in the eigenbasis of $H_0$ \cite{pro}. In the linear
response approximation one then obtains
\begin{equation}\label{l08}
  f_{\epsilon}(\tau)= e^{-\epsilon \, C_{\rm freeze}(\tau)}\,.
\end{equation}
where $C_{\rm freeze}$ differs from the expression (\ref{l05})
derived previously only  by the fact that the term $\tau^2/\beta$ is
missing on the right hand side of the equation \cite{pro}. The
resulting decay of the fidelity amplitude is extremely slow. It will
be discussed below and compared with the exact result as obtained
from the supersymmetry calculation.

\section{The purely imaginary antisymmetric perturbation}

To apply the supersymmetric technique of Reference
\onlinecite{stoe04b} the Hamiltonian needs to be invariant under the
action of the orthogonal/unitary group. This is not the case for a
GOE perturbation with a deleted diagonal. However a purely imaginary
antisymmetric matrix meets with both requirements, zero diagonal
elements and orthogonal symmetry. Therefore it is an ideal
candidate. We consider the Hamiltonian
\begin{equation}\label{a01}
    H_\lambda= H_0+\imath\lambda V\,,
\end{equation}
where $H_0$ is taken from the GOE, i.\,e.
\begin{equation}\label{a02}
    \left<(H_0)_{ij}(H_0)_{kl}\right>_{H_0}\ =\ \frac{N}{\pi^2}
     \left(\delta_{ik}\delta_{jl}+\delta_{il}\delta_{jk}\right)\,,
\end{equation}
and $V$ is real antisymmetric, i.\,e.
\begin{equation}\label{a03}
    \left<V_{ij}V_{kl}\right>_V=
     \left(\delta_{ik}\delta_{jl}-\delta_{il}\delta_{jk}\right)\,.
\end{equation}
The variance of the matrix elements has been chosen to have a mean
level spacing of one for $H_0$ and of order $1/\sqrt{N}$ for $V$.
Note that  in contrast to Reference \onlinecite{stoe04b} the mean
density of states does not remain constant with increasing
perturbation, but decreases with increasing $\lambda$. In fact it is
irrelevant in the present context, whether a defolding to a constant
mean density of states is performed or not. Such a defolding would
imply an additional factor of $1/\sqrt{1+(\pi\lambda)^2/N}$ on the
right hand side of Equation (\ref{a01}), which in the final limit
$N\to\infty$ reduces to one. These definitions are consistent with
the normalization used by Gorin et al.~\cite{gor04}.
Expressing
$f_\epsilon(\tau)$, see Eq.~(\ref{l02}), in terms of its Fourier
transform
\begin{equation}\label{04}
  f_\epsilon(\tau)=\int dE_1\,dE_2
  e^{2\pi\imath\left(E_1-E_2\right)\tau}\,\ree\, ,
\end{equation}
we have
\begin{equation}\label{05}
  \ree\propto\frac{1}{N}\left<{\tr}\left(
  \frac{1}{E_{1-}-H_0-\imath\lambda V}\,\frac{1}{E_{2+}-H_0}\right)\right>\,,
\end{equation}
with $E_\pm=E\pm\imath\eta$. We rewrite $\ree$ using the formula
\begin{equation}\label{06}
\tr\frac{1}{AB}=\left.\sum_{n,m}\frac{\partial}{\partial
J_{nm}}\frac{\partial}{\partial K_{mn}}
\frac{\det(A+J)}{\det(A-J)}\frac{\det(B+K)}{\det(B-K)}\right|_{J=K=0}\ ,
\end{equation}
In our case $A=E_{2+}-H_0$ is real symmetric and
$B=E_{1-}-H_0-\imath\lambda V$ is Hermitean. According to the
universality classes of $A$ and $B$ we write the determinants as
Gaussian integrals over real  and complex wave functions,
respectively. We obtain
\begin{eqnarray}\label{07}
&&\left.\sum_{n,m}\frac{\partial}{\partial
J_{nm}}\frac{\det(A+J)}{\det(A-J)}\right|_{J=0}\ =\ -i\int
d[x]d[y]d[\xi]d[\xi^*]\nonumber\\
&&\qquad\sum_{n,m}(x_nx_m+y_ny_m-\xi_n^*\xi_m-\xi^*_m\xi_n)\nonumber\\
&&\qquad e^{-ix^TAx-iy^TAy-i\xi^\dagger A\xi-i\xi^T A\xi^*},
\end{eqnarray}
where the commuting integration variables are real. We adopt the usual convention
and use latin letters for commuting,
and greek ones for anticommuting variables, respectively. For $B$ we
obtain instead
 \begin{eqnarray}\label{08}
&& \left.\sum_{n,m}\frac{\partial}{\partial
K_{mn}}\frac{\det(B+K)}{\det(B-K)}\right|_{K=0}\ =\ i\int
d[a]d[b]d[\eta]d[\eta^*]\nonumber\\
&&\qquad\sum_{n,m}(a_ma_n+b_mb_n-\eta_m^*\eta_n-\eta_m\eta_n^*)\nonumber\\
&& e^{i(z^\dagger Bz+ 2\eta^\dagger B\eta)},
\end{eqnarray}
where $z_i$ is complex, and $a_i$ and $b_i$ are its real and
imaginary part, respectively.
Collecting the results we arrive at
\begin{widetext}
\begin{eqnarray}\label{06a}
 && \ree\ \propto\ \frac{1}{N}\int d[a]\,d[b]\,d[x]\,d[y]d[\xi]\,d[\xi^*]\,d[\eta]\,d[\eta^*]\ e^{-\imath \left[E_{2+}\sum_n (x_n^2+y_n^2+2\xi^*_n\xi_n)-E_{1-}\sum a(_n^2+b_n^2+2\eta^*_n\eta_n) \right]}\nonumber\\
  &&\quad \sum\limits_{n,m}
  (x_nx_m+y_ny_m-\xi_n^*\xi_m-\xi_n\xi_m^*)(a_ma_n+b_mb_n-\eta_m^*\eta_n-\eta_m\eta_n^*)
  \left\langle e^{\imath \lambda\sum_{n,m} V_{nm}(a_nb_m-a_mb_n-\imath\eta^*_n\eta_m+\imath\eta^*_m\eta_n)}\right\rangle_V\nonumber\\
 &&\qquad\qquad  \left\langle e^{-\imath \sum_{n,m}{H_0}_{nm}\left(x_nx_m+y_ny_m-a_na_m-b_nb_m+\xi_n^*\xi_m+\xi^*_m\xi_n-
   \eta_n^*\eta_m-\eta_m^*\eta_n\right)}\right\rangle_{H_0}\ .
\end{eqnarray}
\end{widetext}
The
commuting integration variables are all real. Now the average is
taken over real symmetric $H_0$ using Eq.~(\ref{a02})
\begin{equation}\label{09a}
\left\langle\dots\right\rangle_{H_0}=\exp\left(-\frac{N}{\pi^2}\Str(LZ)^2\right)
\end{equation}
where $Z$ is a supermatrix given by   $Z=$ $\sum_n\zb_n\zbd_n$
with $\zb_n^T$ $=(x_n,y_n,\xi_n,\xi^*_n,a_n,b_n,\eta_n,\eta^*_n)$,
and $L=\diag(1_4,-1_4)$ in the advanced--retarded block notation.
$Z$ is exactly the matrix given in Table 4.1 of Reference
\onlinecite{ver85a}, denoted by VWZ in the following. The matrix
$B$ has an orthosymplectic symmetry, i.~e. in Boson--Fermion block
notation the Boson-Boson block is real symmetric and the
Fermion--Fermion block is Hermitean selfdual. For the $V$ average
we obtain with Eq.~(\ref{a03})
\begin{equation}\label{10}
\left\langle\dots\right\rangle_{V}=\exp\left(-\lambda^2\Str({\bf
K}T)^2\right)\ .
\end{equation}
Here $T=\sum_n \ab_n\ab^\dagger_n$ with $\ab^T_n=
(a_n,b_n,\eta_n,\eta_n^*)$ the supermatrix ${\bf K}$ is given by
\begin{equation}\label{11}
{\bf K}=\left(\begin{matrix}{-\sigma_y&0\cr
                         0&\sigma_z}\end{matrix}\right)\,,
\end{equation}
where we used
the Pauli matrices
\begin{equation}\label{12}
 \sigma_x=\left(\begin{array}{cc}
    0 & 1 \\
    1 & 0
  \end{array}\right)\ ,\
   \sigma_y=\left(\begin{array}{cc}
    0 & -\imath \\
    \imath & 0
  \end{array}\right)\,,\
  \sigma_z=\left(\begin{array}{cc}
    1 & 0 \\
    0 & -1
  \end{array}\right).
\end{equation}
In ${\bf K}$ the off--diagonal nature of the perturbation is
encoded.

The subsequent steps are the same as described in
\cite{ver85a,stoe04b}. After transforming equations (\ref{09a})
and (\ref{10}) by means of two Hubbard-Stratonovich
transformations, the integrations over the the $a,b,x,y$
variables, and over the auxiliary variables of one
Hubbard-Stratonovich transformation can be performed resulting in
\begin{eqnarray}\label{21}
 && \ree\ \propto\ \frac{\pi^4}{4N^3}
  \int d[\sigma]\Str\left({\bf P}\sigma_{RA}^\dagger{\bf
  P}\sigma_{RA}\right)\nonumber\\
  &&\quad e^{-\frac{\pi^2}{4N}\Str \sigma^2}
 e^{\frac{\pi^4}{4N^2}\lambda^2\Str\left({\bf K}\sigma_{RR} \right)^2}\nonumber\\
 && \quad\left[\Sdet\left(\begin{array}{cc}
    \sigma_{AA}-E_{1-}& \sigma_{RA}^\dagger \\
    \sigma_{RA} & \sigma_{RR}-E_{2+} \
  \end{array}\right)\right]^{-N/2}\,.
\end{eqnarray}
Here ${\bf P}=\diag(1,1,-1,-1)$. The matrix $\sigma$ has the same
orthosymplectic symmetry as $B$ and reads in
advanced-retarded block notation
\begin{eqnarray}\label{13}
\sigma&=&\left[\begin{matrix}{\sigma_{AA}&\sigma_{RA}^\dagger\cr
                             \sigma_{RA}&\sigma_{RR}}\end{matrix}\right]\,
                 \quad \ .
\end{eqnarray}
Introducing the notation $E_{1/2}=\bar{E}\pm E/2$, and
substituting $\sigma_{AA}$ and $\sigma_{RR}$ by $\sigma_{AA}+E/2$
and  $\sigma_{RR}-E/2$, respectively, we obtain
 \begin{eqnarray}\label{23}
   \ree&\propto&\frac{\pi^4}{4N^3}
   \int d[\sigma]\Str\left({\bf P}\sigma_{RA}^\dagger{\bf P}\sigma_{RA}\right)\nonumber\\
   && e^{-\frac{\pi^2}{4N}\left[\sigma^2 + E\ \Str(\sigma_{AA}-\sigma_{RR})\right]}\nonumber\\
   && e^{\frac{\pi^4\lambda^2}{4N^2}\left[\Str\left({\bf K}\sigma_{RR} \right)^2-E\Str{\bf K}\sigma_{RR}\right]}
   \nonumber\\&\times&\left[\Sdet(\sigma-\bar{E})\right]^{-N/2}.
 \end{eqnarray}

This expression can be evaluated in the limit $N\to \infty$ by a
saddle point approximation.

\section{Saddle point Approximation}

The next steps are a direct repetition of the corresponding ones
in Reference \onlinecite{stoe04b}. We shall adopt the notation
from VWZ, which is the main source for the following calculations.
First we diagonalize $\sigma$,
\begin{equation}\label{24}
  \sigma=T_0^{-1}R^{-1}\sigma_DRT_0\,,
\end{equation}
where $R$ is block diagonal  and
\begin{equation}\label{25}
  T_0=\left(\begin{array}{cc}
    \sqrt{1+\tb\tc} & \imath\tb \\
    -\imath\tc & \sqrt{1+\tc\tb} \
  \end{array}\right)\,
\end{equation}
(see VWZ, Eqs. (5.28+29)). The integration of the diagonal variables
of $\sigma$ can be performed by means of the saddle point
approximation. $\sigma_D$ at the saddle point reads
\begin{equation}\label{27}
  \sigma_D=\left(\begin{array}{cc}
    s_A & {\bf 0} \\
    {\bf0} & s_R \
  \end{array}\right)\,,
\end{equation}
where the advanced and retarded saddle points are given by
\begin{eqnarray}\label{28}
  s_{A/R}&=&\frac{1}{2}\left(\bar{E}\pm\imath\Delta\right)\,,\nonumber\\
  \Delta&=&\frac{2N}{\pi}\sqrt{1-\left(\frac{\pi\bar{E}}{2 N}\right)^2}\ =\ \frac{2N}{\pi}\rho\,,
\end{eqnarray}
and $\rho$ is the density of states. In the following we shall
restrict ourselves to the band centre, $\bar{E}=0$, where Eq.
(\ref{28}) reduces to $s_{A/R}=\pm\imath\pi/N$. We then have for
the matrix $\sigma$ at the saddle point
\begin{equation}\label{29}
  \sigma\ =\ \frac{N}{\pi}\left(\begin{array}{cc}
    \imath\left({\bf 1}+2\tb\tc\right) & 2\tb\sqrt{{\bf 1}+\tc\tb}\\
     2\tc\sqrt{{\bf 1}+\tb\tc} & \imath\left(-{\bf 1}-2\tc\tb\right)
    \end{array}\right)\,.
\end{equation}
The matrix $R$ (see Eq. (\ref{24})) does not enter, since it
commutes with $\sigma_D$ at the saddle point.  We obtain

\begin{eqnarray}\label{26}
   && \ree\ \propto\ \frac{\pi^4}{4N^3}\int {\mathcal
  F}(t_{12})d[t_{12}]\Str\left({\bf P}\sigma_{AR}{\bf P}\sigma_{RA}\right)\nonumber\\
  &&\qquad e^{-\frac{\pi^2}{4N}\left[E\Str(\sigma_{AA}-\sigma_{BB})-\frac{\pi^2\lambda^2}{N}
  \Str \left({\bf K}\sigma_{AA} \right)^2\right]}\, ,
\end{eqnarray}

where the integral is over the elements of the matrix $t_{12}$,
parametrizing the saddle-point manifold. The function ${\cal
F}(t_{12})=$ $\Sdet^{-1/2}(1+t_{12}t_{21})$ is the Berezinian of the
coordinate transformation Eq.~(\ref{25}). Using Eq.~(\ref{29}) the
various terms entering Eq.~(\ref{26}) may be written as
\begin{eqnarray}\label{30a}
  \Str \left(\sigma_{AA}-\sigma_{RR}\right)&=&\imath\frac{4N}{\pi}\Str\left(\tb\tc\right)
  \,,\\\label{30b}
  \Str \left({\bf K}\sigma_{RR} \right)^2&=&\left(\frac{\imath N}{\pi}\right)^2\Str\left[{\bf K}\left({\bf 1}+2\tb\tc\right)\right]^2
  \\\label{30c}
  \Str(\sigma_{AR}{\bf P} \sigma_{RA}{\bf P})&=&
 \left(\frac{2N}{\pi}\right)^2
 \Str\left(\tc\sqrt{{\bf 1}+\tb\tc}{\bf P}\right.\nonumber\\
 &&\qquad\quad\times\left.\tb\sqrt{{\bf 1}+\tc\tb}
  {\bf P}\right)
  \,.
\end{eqnarray}
We proceed further by diagonalizing the matrices $\tb$ and $\tc$.
This is achieved by the radial decomposition
 \begin{equation}\label{31a}
  \tb=U_1^{-1}
    M  U U_2\,,\qquad
  \tc=U_2^{-1}
   U^\dag M
   U_1\,,
\end{equation}
with diagonal $M=\diag(\mu_1,\mu_2,\imath\mu,\imath\mu)$. The matrix
$U=1_2\oplus \widehat{U}$ is a $4\times 4$ block diagonal
matrix, with $\widehat{U}$ $\in {\rm SU}(2)$ (see VWZ, Eq.
(I.18)). The $U_i$ $(i=1,2)$ may be parameterised as $U_i=V_iO_i$,
where the $O_i=\widehat{O}_i\oplus 1_2$ are $4\times 4$ block
diagonal, with $\widehat{O}_i\in {\rm SO}(2)$. The
parameterisation of the $V_i$ in terms of anti-commuting variables
is postponed to App.~\ref{appA}. If we moreover introduce
\begin{eqnarray}\label{36}
  X=M^2=\diag (x,y,-z,-z)\,,\nonumber\\\qquad x=\mu_1^2\,,\quad y=\mu_2^2\,,\quad z=\mu^2\, ,
\end{eqnarray}
we can write Eqs.~(\ref{30a}) to (\ref{30c}),
 using Eq.~(\ref{31a}) and Eq.~(\ref{36})

\begin{eqnarray}\label{33a}
  \Str \left(\sigma_{AA}-\sigma_{RR}\right)&=&\imath\frac{4N}{\pi}\Str X
  \,,\\\label{33b}
  \Str \left({\bf K}\sigma_{RR} \right)^2&=&\left(\frac{\imath N}{\pi}\right)^2\Str\left[{\bf K_1}\left({\bf 1}+2X\right)\right]^2
  ,\\\label{33c}
  \Str(\sigma_{AR}{\bf P} \sigma_{RA}{\bf P})&=&
  \left(\frac{2N}{\pi}\right)^2\Str\left(\sqrt{X}\sqrt{{\bf 1}+X}{\bf P_1}\right.\nonumber\\
  &&\qquad\quad\times\left.\sqrt{X}\sqrt{{\bf 1}+X}
  {\bf P_2}\right)
  \,,
\end{eqnarray}

where
\begin{equation}\label{34}
{\bf K_1}=U_1{\bf K}U_1^{-1}\,,\,{\bf P_1}=U_1{\bf
P}U_1^{-1}\,,\,{\bf P_2}=UU_2{\bf P}U_2^{-1}U^{-1}\,.
\end{equation}
Under the transformations Eq.~(\ref{31a}) and Eq.~(\ref{36}) the
measure transforms as
\begin{equation}
\label{mea} d[t_{12}]= {\mathcal G}(X)
d\mu(U_1)d\mu(U_2)d\mu(U)d[X]\ .
\end{equation}
The function ${\mathcal G}$ has been calculated in VWZ (Eq.~K.17). The average in Eq. (\ref{26}) is over the elements of the
matrices $X,U,U_1,U_2$. Only ${\bf P_2}$ depends on the matrix
elements of $U_2$. It will be shown in App.~\ref{appA} that $U_2{\bf
P}U_2^{-1}$ averaged over the matrix elements of $U_2$ is nothing
but a multiple of the four-dimensional unit matrix. Thus the $U$
dependence cancels. We are then left with an average over $x, y, z$,
and the matrix elements of $U_1$. Inserting these results into Eq.
(\ref{26}), we get

\begin{eqnarray}\label{35}
  &&  \ree\ \propto\ \frac{1}{N}\int {\mathcal F}(X)
    {\mathcal G}(X)d[X]
     d\mu(U_1)\\
  &&\quad \Str\left[X(X+{\bf1}){\bf P_1}\right]
  e^{-\pi\imath E\Str X-\frac{\epsilon}{16} \Str \left[({\bf1}+2X){\bf K_1}\right]^2} \ ,\nonumber
\end{eqnarray}

where we employed the definition  $\epsilon=4\pi^2\lambda^2$, see
above. The Berezinians ${\mathcal F}(X)$ and ${\mathcal G}(X)$ can
be comprised in one measure function $\mu(X)$ which was given in
VWZ:
\begin{equation}
\mu(X)=\frac{|x-y|}{\sqrt{xy(x+1)(y+1)}}
  \frac{z(1-z)}{(z-x)^2(z-y)^2}.
\end{equation}
Substituting expression (\ref{35}) for $\ree$ into Eq. (\ref{04}),
and introducing $E=(E_1-E_2)/2$ and $\bar{E}=(E_1+E_1)/2$ as new
integration variables, the $E$ integration generates a delta
function, whereas the $\bar{E}$ integration corresponds to an
energy average. The result is (see   Reference
\onlinecite{stoe04b} for details)

\begin{eqnarray}\label{36a}
  && f_\epsilon(\tau)\propto\frac{1}{N}\int_0^\infty du\int_0^\infty dv\int_0^1 dz\mu(u,v,z)d\mu(U_1)\nonumber\\
  &&\qquad\quad\delta\left(\tau-u-z\right)\Str(X(X+{\bf1}){\bf P_1})\nonumber\\
  &&\qquad\quad\exp\left(-\frac{\epsilon}{16} \Str \left[({\bf1}+2X){\bf K_1}\right]^2\right)\ ,
\end{eqnarray}
 with $u=(x+y)/2$. In addition we shall use
$v=(x-y)/2$ as another new variable, and replace $z$ by $\tau-u$
everywhere, which is admissible because of the presence of the
delta function. The integration domains of the radial variables $u,v$ and $z$ are dictated by the hyperbolic symmetry
of the saddle point manifold, i.~e. we have non--compact integration domains $0< x,y<\infty$ for the bosonic coordinates
$x,y$ and a compact integration domain $0<z<1$ for the fermionic coordinate $z$. For more detail on this point, see VWZ.
We still have to integrate over the matrix
elements of $U_1$.

\section{Integration over the Grassmann variables}

We recall that $U_1=V_1O_1$. Since $O_1$ commutes with ${\bf P}$
and with ${\bf K}$, the $O_1$ integration is trivial, and we are
left with the integration over $V_1$. The parametrization of $V_1$
in terms of Grassmannian variables and the
 calculation of the traces in equation (\ref{36a}) is quite
involved, and is postponed to the appendices. Here we note only
the results:
\begin{eqnarray}\label{55}
 && \Str\left[X(X+{\bf1}){\bf P_1}\right]
  \ =\ 4v(2u+1)B\nonumber\\
  &&\qquad\qquad+2[2u(u+1)-\tau(2u+1-\tau)+v^2]\nonumber\\
  &&\qquad\qquad+4[\tau(2u+1-\tau)+v^2](A-2\bar{a}),
\end{eqnarray}
and
\begin{eqnarray}\label{67a}
  &&\frac{1}{8}\Str \left[({\bf1}+2X){\bf
    K_1}\right]^2=\nonumber\\
    &&\qquad\ \tau(2u+1-\tau)-v^2+2(A-D)(\tau^2-v^2),
\end{eqnarray}
where
\begin{equation}\label{47a}
A=\alpha\alpha^*+\beta\beta^*\,,\quad
B=\alpha\alpha^*-\beta\beta^*\,,\quad
D=\imath(\alpha\beta^*-\beta\alpha^*)\,.
\end{equation}
and $\quad\bar{a}=\alpha\alpha^*\beta\beta^*$. $\alpha$,
$\alpha^*$, $\beta$, $\beta^*$ are anticommuting variables. It
follows
\begin{eqnarray}\label{68}
 e^{-\frac{\epsilon}{16} \Str \left[({\bf1}+2X){\bf
    K_1}\right]^2} &=&
    e^{-\frac{\epsilon}{2}
    \left[\tau(2u+1-\tau)-v^2+2(A+D)(\tau^2-v^2)\right]}\nonumber\\
    &=&[1-\epsilon(A-D)(\tau^2-v^2)]\nonumber\\
    &&\quad e^{-\frac{\epsilon}{2}\left[\tau(2u+1-\tau)-v^2\right]} \ .
\end{eqnarray}
These results are inserted into equation  (\ref{36a}). The measure
is given by
\begin{equation}
\label{meas} d\mu(U_1)\ =\ 2\pi d\mu(V_1)\ \propto \  d\alpha
d\alpha^* d\beta d\beta^* \ .
\end{equation}
Therefore only the terms proportional to $\bar{a}$ survive the
integration over the antisymmetric variables. We obtain
\begin{eqnarray}\label{71}
    &&f_\epsilon(\tau)\ \propto \ \frac{1}{N}\int \mu(u,v,z)\delta\left(\tau-u-z\right)
    e^{-\frac{\epsilon}{2}\left(\tau(2u+1-\tau)-v^2\right)}\nonumber\\
    &&\qquad [1+\epsilon(\tau^2-v^2)][\tau(2u+1-\tau)+v^2]dudvdz \ ,
\end{eqnarray}
which is almost our final result.

\section{Result and discussion}
The final result is obtained by an VWZ-like integral (see VWZ, Eq.
(8.10)) and is given in the present case by

\begin{widetext}
\begin{eqnarray}\label{09}
  f_\epsilon(\tau)&=&2\int\limits_{{\rm Max}(0,\tau-1)}^\tau du\int\limits_0^u
  \frac{v\,dv}{\sqrt{[u^2-v^2][(u+1)^2-v^2]}}
  \frac{(\tau-u)(1-\tau+u)}{(v^2-\tau^2)^2}
  \nonumber\\
  &&
  \times [1+\epsilon(\tau^2-v^2)][\tau(2u+1-\tau)+v^2]e^{-\frac{\epsilon}{2}[\tau(2u+1-\tau)-v^2]}\,.
\end{eqnarray}
\end{widetext}

The constant of proportionality was fixed by the condition
$f_\epsilon(0)=1$ (see Reference \onlinecite{stoe05}). The only
difference of Eq.~(\ref{09}) to Reference \onlinecite{stoe05},
where a GOE perturbation was considered, is the additional factor
$[1+\epsilon(\tau^2-v^2)]$ in the integrand, and a minus sign with
the $v^2$ term in the exponent, where in the GOE case there is a
plus sign.

\begin{figure}
   \includegraphics[width=\columnwidth]{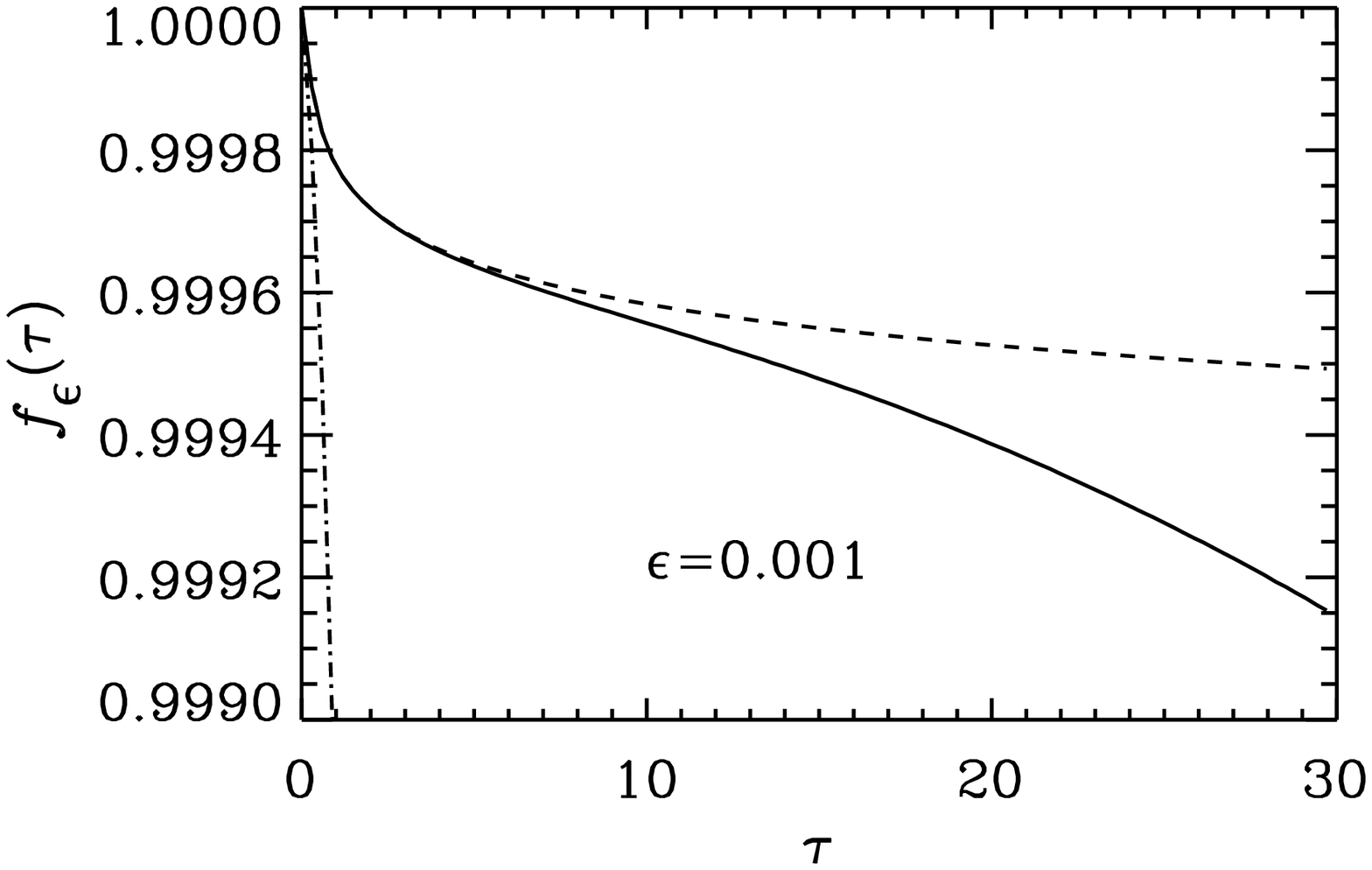}\\
\includegraphics[width=\columnwidth]{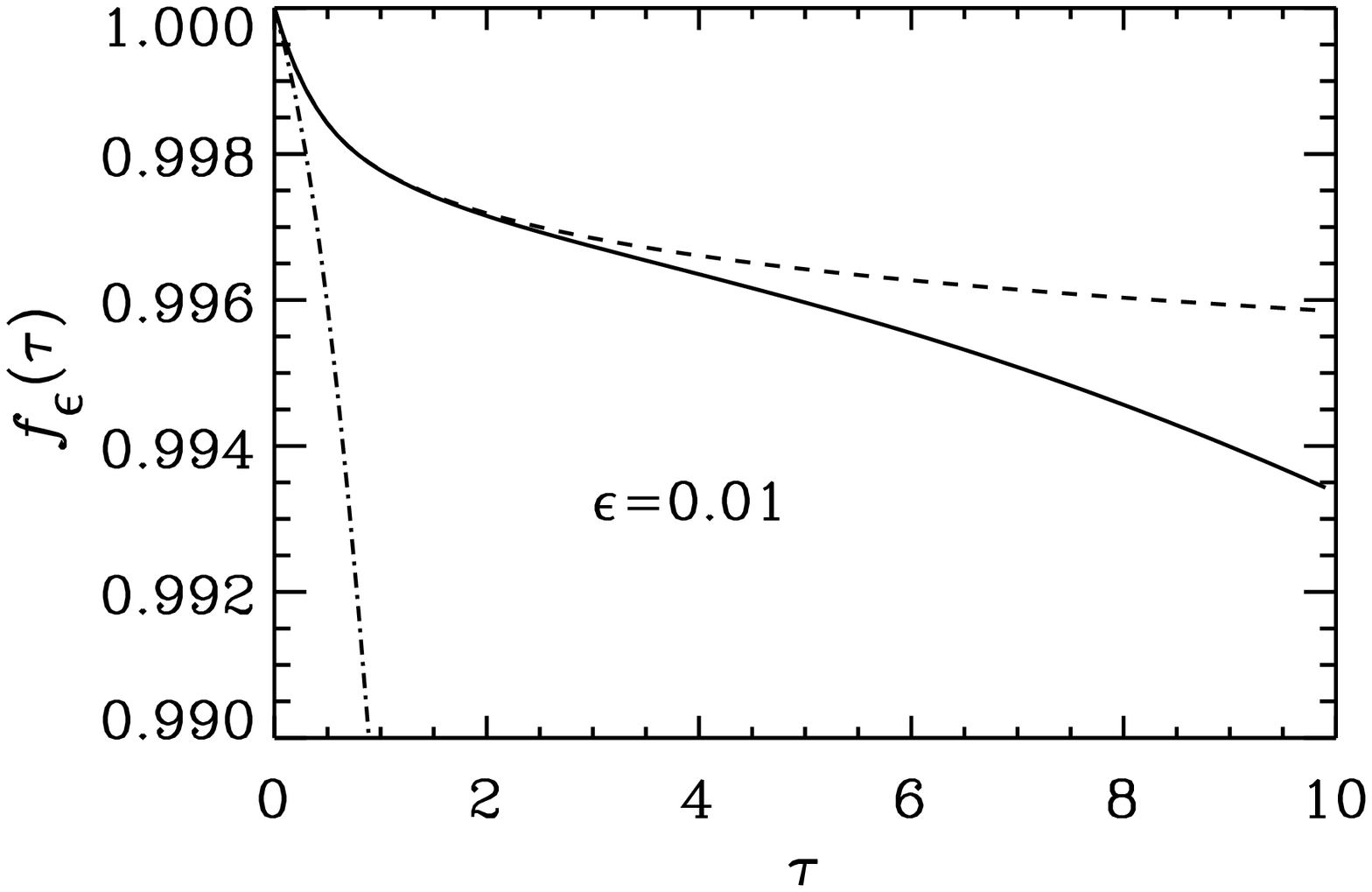}\\
\includegraphics[width=\columnwidth]{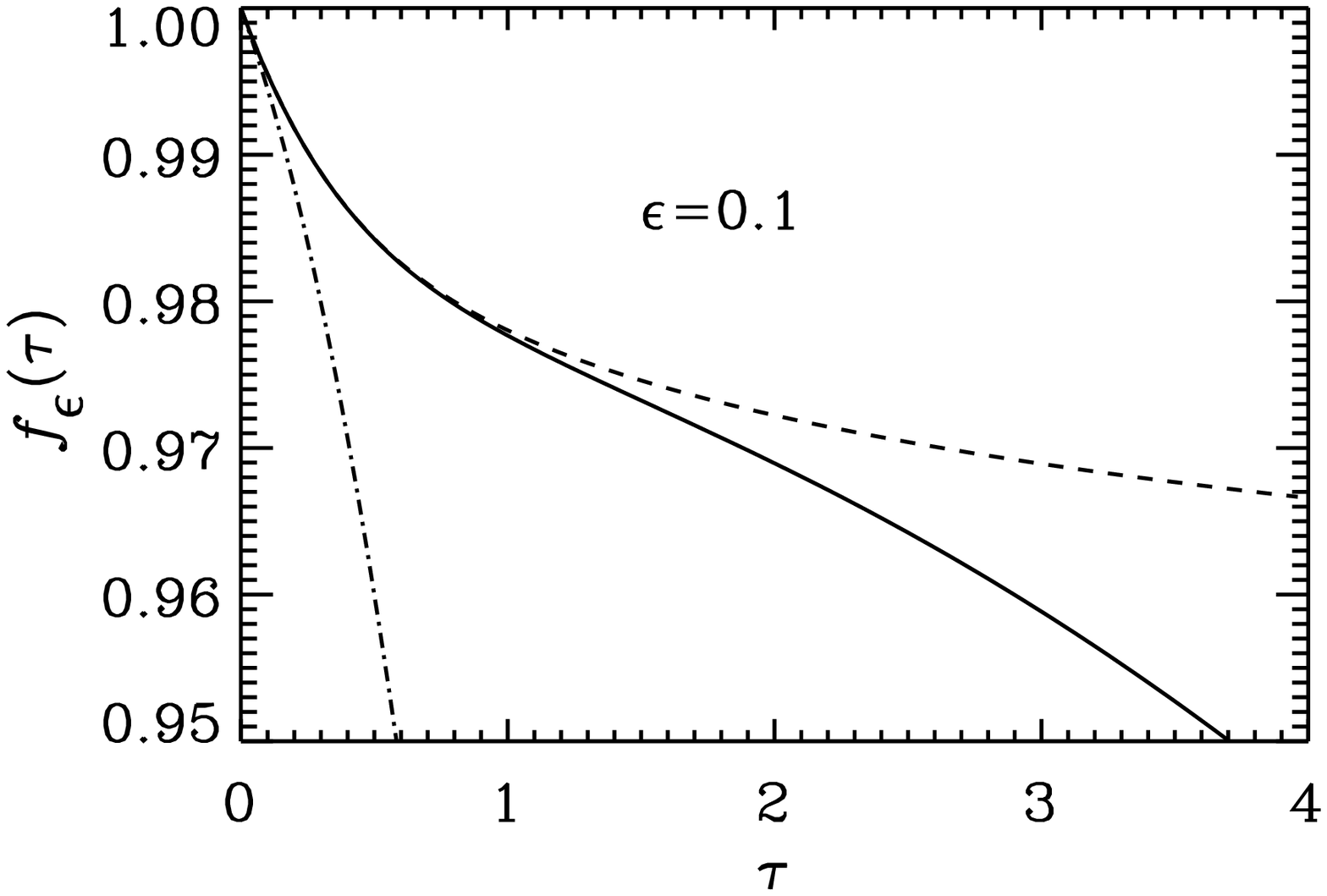}\\
  \caption{Ensemble average of the fidelity amplitude $f_\epsilon(\tau)$ with $H_0$ taken from the GOE and a
  purely imaginary antisymmetric perturbation (solid line, calculated from Eq. (\ref{09}))
  for different perturbation strengths $\epsilon$.
  For comparison the result
  from the linear response approximation (dashed line), and for a GOE
  perturbation (dashed-dotted line) are shown as well. }
\label{fig:freeze}
\end{figure}
Figure \ref{fig:freeze} shows the fidelity decay for different
perturbations, as calculated from Eq. (\ref{09}), together with
the result from the exponentiated linear response approximation.
For comparison the fidelity decay for the case of a GOE
perturbation \cite{stoe04b}) is shown as well. We see that the
linear response approximation is able to describe the fidelity
decay for quite a long time very well. For still larger times the
linear response approximation underestimates the decay, as
compared to the exact result, but still the decay is by orders of
magnitudes slower as for a GOE perturbation.

\begin{figure}
    \includegraphics[width=\linewidth]{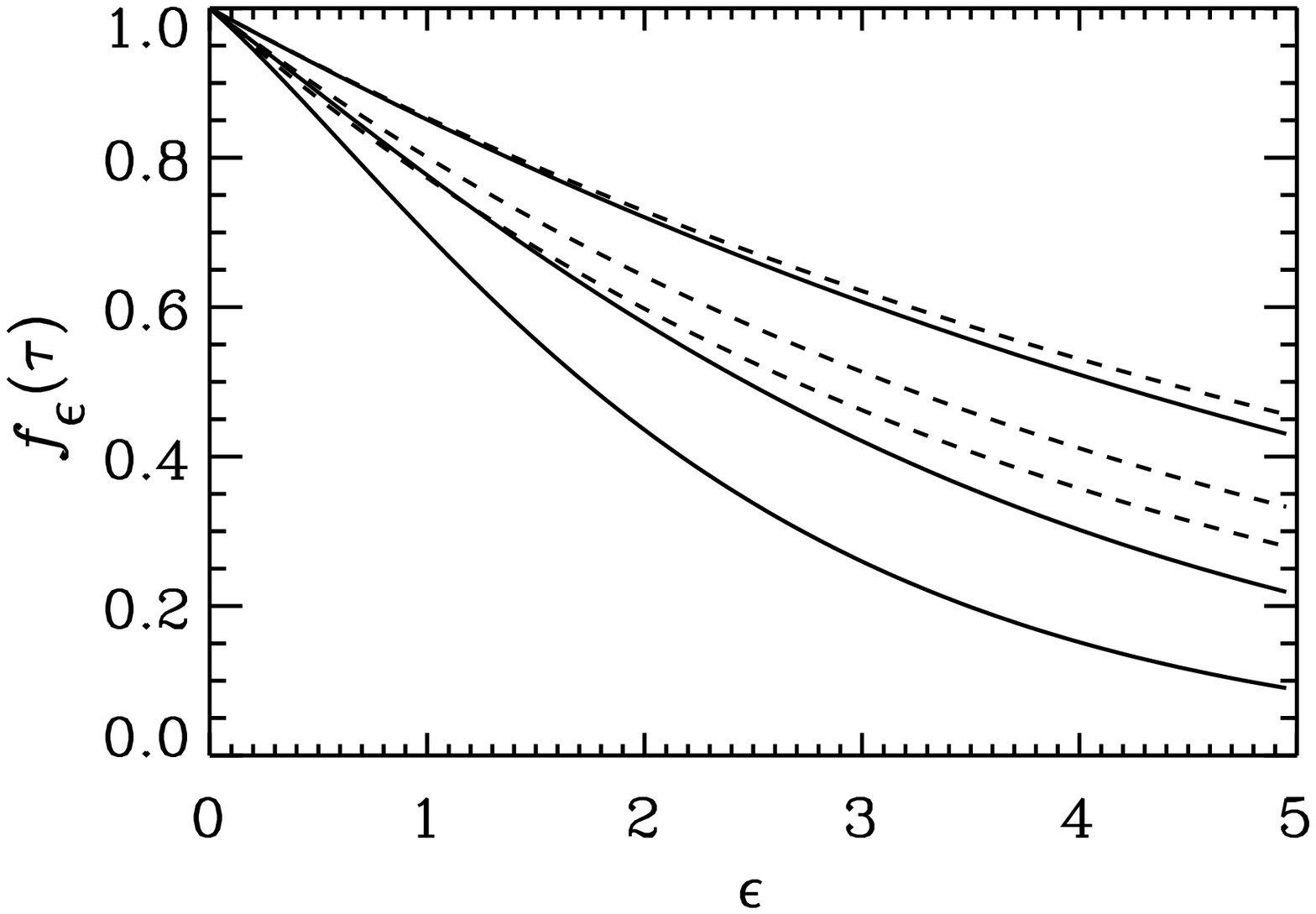}\\
  \caption{Ensemble average of the fidelity amplitude $f_\epsilon(\tau)$
 for  an imaginary antisymmetric perturbation  as a function of $\epsilon$
 for three fixed values of $\tau=
 0.5,1.0,1.5$ (from top to bottom, solid lines).
Again the results  from the linear response approximation are
shown for comparison (dashed lines).}\label{fig:freeheis}
\end{figure}

Figure \ref{fig:freeheis} shows the fidelity $f_\epsilon(\tau)$
for three fixed values 0.5, 1.0, 1.5 of $\tau$ as a function of
the perturbation $\epsilon$. The figure demonstrates that the
freezing effect not unexpectedly becomes less and less pronounced
with increasing perturbation, though the decay is always by orders
of magnitude slower than for the case of a GOE perturbation (not
shown). It is further seen that the linear response approximation
works very well up to about half the Heisenberg time, but
underestimates the decay more and more for increasing $\tau$
values.

We thus can conclude that the fidelity freeze is not an artefact
of the linear response approximation but is also present in the
exact calculation. Due to the perfect and well established
correspondence between random matrices and chaotic quantum systems
\cite{cas80,boh84b,guh98} this result provides an important new
mechanism of preserving quantum stability.\\

\begin{acknowledgments}
The motivation to this work came from numerous discussions with
T.~Gorin, T.~Prosen, T.~Seligman, and M. {\v Z}nidari{\v c}, as well
as from microwave experiments on the subject of fidelity, which had
been performed by R.~Sch\"afer together with one of the authors
(H.-J. St.). The experiments had been supported by the Deutsche
Forschungsgemeinschaft.
\end{acknowledgments}

\appendix

\section{Calculation of ${\bf Str}\left[X(X+{\bf1}){\bf P_1}\right]$}
\label{appA}

In terms of Pauli matrices $X$ may be expressed as
\begin{equation}\label{38}
  X=-z{\bf 1}+\left(\begin{array}{cc}
    \hat{X} & {\bf0} \\
    {\bf0} & {\bf0}
  \end{array}\right)\,,
\end{equation}
where
\begin{equation}\label{39}
  \hat{X}=\left(\begin{array}{cc}
    x+z & 0 \\
    0 & y+z
  \end{array}\right)
  =\tau{\bf 1}+v\sigma_z\,.
 \end{equation}
It follows
\begin{eqnarray}\label{40}
&&\Str\left[X(X+{\bf1}){\bf P_1}\right]=4z(z-1)\nonumber\\
&&+(1-2z)\tr\left[\hat{X}\left({\bf P_1}\right)_{\rm u.l.}\right]
  +\tr\left[\hat{X}^2\left({\bf P_1}\right)_{\rm u.l.}\right]
\end{eqnarray}
where it was used that $\Str{\bf P_1}=\Str{\bf P}=4$, and where
$\left({\bf P_1} \right)_{\rm u.l.}$ denotes the upper left
submatrix of ${\bf P_1}$. As was already mentioned, matrices $U_1$
and $U_2$ entering the calculation of $\bf P_1$ and $\bf P_2$ (see
Eq. (\ref{34})) are parameterized as
\begin{equation}\label{41}
  U_p=V_pO_p\,,\qquad (p=1,2)
\end{equation}
where
\begin{equation}\label{42}
  O_p=\left(\begin{array}{cc}
    \hat{O}_p & {\bf0} \\
    {\bf0} & {\bf 1}
  \end{array}\right)\, ,
\end{equation}
and $\hat{O}_1$ and $\hat{O}_2$ are $2\times2$ orthogonal matrices
(see VWZ, Eq. (I.13)). The matrices $V_p$ may be parametrized as
(see VWZ, Eq. (K.26))
\begin{equation}\label{43}
  \left(V_p\right)^{\pm1}=1\pm\imath^{p-1}Y_p+\frac{1}{2}\imath^{2(p-1)}Y_p^2\pm\frac{1}{2}\imath^{3(p-1)}Y_p^3
  +\frac{3}{8}Y_p^4
\end{equation}
where matrices $Y_1$ and $Y_2$ are given by
\begin{equation}\label{44}
Y_p=\left(\begin{array}{cc}
  {\bf0} & - \zeta_p^\dag \\
  \zeta_p & {\bf0}
\end{array}\right)
\end{equation}
where
\begin{equation}\label{45}
  \zeta_p=\left(\begin{array}{cc}
    \alpha_p & \beta_p \\
    \alpha_p^* & \beta_p^*
    \end{array}\right)\,,\qquad
    \zeta_p^\dag=\left(\begin{array}{cc}
    \alpha_p^* & -\alpha_p \\
    \beta_p^* & -\beta_p
  \end{array}\right)
\end{equation}
(see VWZ, Eqs. (K.23 + 25)). Note the convention
$(\alpha^*)^*=-\alpha$ for antisymmetric variables. In VWZ, Eq.
(I.13) the sequence of the matrices on the right hand side of Eq.
(\ref{41}) is reversed. Both parameterizations are equivalent and
can be transformed into each other by a straightforward
transformation of the $\alpha_p, \beta_p$ variables.

We are now going to calculate ${\bf P_1}=U_1{\bf
P}U_1^{-1}=V_1O_1{\bf P}O_1^{-1}V_1^{-1}$.  To simplify notations,
we shall omit the lower index `1'  in the following. The
calculation for ${\bf P_2}$ proceeds in the very same way. Since
${\bf P}=\diag(1,1,-1,-1)$ commutes with $O$, we are left with
\begin{equation}\label{46}
  {\bf P_1}=V{\bf P} V^{-1}=V\left(\begin{array}{cc}
    {\bf 1} & {\bf0} \\
    {\bf0} & {\bf -1}
  \end{array}\right)V^{-1}\,.
\end{equation}

For the further calculation it is suitable to introduce the
quantities
\begin{eqnarray}\label{47}
A=\alpha\alpha^*+\beta\beta^*\,,\quad
&B&=\alpha\alpha^*-\beta\beta^*\,,\nonumber\\
C=\alpha\beta^*+\beta\alpha^*\,,\quad
&D&=\imath(\alpha\beta^*-\beta\alpha^*)\,.
\end{eqnarray}
$A$, $B$, $C$ obey the relations
\begin{equation}\label{48}
  A^2=2\bar{a}\,,\quad B^2=-2\bar{a}\,,\quad C^2=-2\bar{a}\,,\quad D^2=-2\bar{a}\,,
  \end{equation}
where $\quad\bar{a}=\alpha\alpha^*\beta\beta^*$, and
\begin{equation}\label{49}
    AB=AC=AD=BC=BD=CD=0\,.
\end{equation}
It follows
\begin{equation}\label{50}
  \zeta^\dag\zeta=-A{\bf1}-B\sigma_z-C\sigma_x\,,\qquad\zeta\zeta^\dag=A {\bf 1}
\end{equation}

As a direct consequence we have
\begin{eqnarray}\label{51}
  Y^2&=&\left(\begin{array}{cc}
    -\zeta^\dag\zeta & {\bf0} \\
    {\bf0} & -\zeta\zeta^\dag
  \end{array}\right)\nonumber\\
  &=&
  \left(\begin{array}{cc}
    A{\bf1}+B\sigma_z+C\sigma_x & {\bf0} \\
    {\bf0} & -A{\bf1}
  \end{array}\right)\,,\nonumber\\ Y^3&=&-AY\,.
\end{eqnarray}

It follows from Eq. (\ref{43})
\begin{eqnarray}\label{52}
  V^{\pm1}&=& 1+\left(\frac{1}{2}-\frac{3}{8}A\right)Y^2\pm\left(1-\frac{A}{2}\right)Y\nonumber\\
  &&=\left(
\begin{array}{cc}
  w & \mp\omega^* \\
  \pm\omega & \bar{w}
\end{array}\right)\,,
\end{eqnarray}
where
\begin{eqnarray}\label{53}
  w&=&\left(1+\frac{A}{2}-\frac{3}{4}\bar{a}\right)+\frac{B}{2}\sigma_z+\frac{C}{2}\sigma_x\nonumber\\
  \omega&=&\left(1-\frac{A}{2}\right)\zeta\,,\qquad
  \omega^\dag=\left(1-\frac{A}{2}\right)\zeta^\dag\nonumber\\
  \bar{w}&=&1-\frac{A}{2}+\frac{3}{4}\bar{a}
\end{eqnarray}

Inserting the results into Eq. (\ref{46}) we have

\begin{widetext}

\begin{equation}\label{54}
  {\bf P_1}=\left(\begin{array}{cc}
    \left(1-4\bar{a}+2A\right){\bf1}+2\left(B\sigma_z+C\sigma_x\right) & 2(1-A)\zeta^\dag \\
    2(1-A)\zeta & (-1-4\bar{a}+2A){\bf1}
  \end{array}\right)\,.
\end{equation}

\end{widetext}

 A corresponding expression is obtained for $\bf P_2$. In the
average over the antisymmetric variables only the $\bar{a}$ terms
survive, i.\,e. $\langle {\bf P_1}\rangle=\langle {\bf
P_2}\rangle\propto {\bf1}$, as was stated above.

Inserting finally the upper left corner element of ${\bf P_1}$
into Eq. (\ref{40}), we end up with Eq. (\ref{55}).

\section{Calculation of ${\bf Str} \left[({\bf1}+2X){\bf K_1}\right]^2/8$}

It is suitable to write

\begin{eqnarray}\label{56}
  \frac{1}{8}\Str \left[({\bf1}+2X){\bf K_1}\right]^2&=&
  \frac{1}{16}\Str \left[({\bf1}+2X),{\bf K_1}\right]^2 \nonumber\\
  &&+\frac{1}{8}
  \Str \left[({\bf1}+2X)^2{\bf K_1}^2\right]\nonumber\\
&=&\frac{1}{4}\Str \left[X,{\bf K_1}\right]^2\nonumber\\
&& +\frac{1}{2}
  \Str \left[X(X+{\bf 1})\right]
\end{eqnarray}

where ${\bf K_1}^2={\bf K}^2={\bf 1}$ was used. The second term on
the right hand side is easily evaluated:
\begin{eqnarray}\label{57}
  \frac{1}{2}\Str \left[X(X+{\bf 1})\right]&=&
 \frac{1}{2}\left[ x(x+1)+y(y+1)+2z(1-z)\right]\nonumber\\
 &=&u(u+1)+v^2+z(1-z)\nonumber\\
 &=&-\tau^2+(2u+1)\tau+v^2\,.
\end{eqnarray}
For the first term on the right hand side we need an expression
for ${\bf K_1}$. Using ${\bf K_1}=U_1{\bf K}U_1^{-1}$ (see Eq.
(\ref{34})) and $U=VO$ (see Eq. (\ref{41})) we may write
\begin{equation}\label{58}
  {\bf K_1}=V_1O_1{\bf K}O_1^{-1}V_1^{-1}=V{\bf K}V^{-1}\,,
\end{equation}
since $K$ (see equation (\ref{11})) commutes with $O$. Using Eq.
(\ref{52}), we obtain
\begin{equation}\label{61}
  {\bf K_1}=\left(\begin{array}{cc}
    k & \kappa^\dag \\
    \kappa &
    \bar{k}
  \end{array}\right)\,,
\end{equation}
where
\begin{eqnarray}\label{62}
  k&=&-w\sigma_y w+\omega^\dag\sigma_z\omega\,,\nonumber\\
  \kappa^\dag&=&-w\sigma_y \omega^\dag-\omega^\dag\sigma_z\bar{w}\,,\qquad
  \kappa=-\omega\sigma_y w-\bar{w}\sigma_z \omega\,,\nonumber\\
  \bar{k}&=&-\omega\sigma_y\omega^\dag+\bar{w}\sigma_z\bar{w}\,.
\end{eqnarray}

Since ${\bf K_1}^2={\bf 1}$, we have
\begin{eqnarray}\label{64}
  \frac{1}{4}\Str \left[X,{\bf K_1}\right]^2&=&\frac{1}{2}\left[
  \Str\left(X{\bf K_1}\right)^2- \Str X^2\right]
  \nonumber\\
    &=&\frac{1}{2}\left[\Str\left(\hat{X}k\right)^2-\Str\hat{X}^2\right]\,,
\end{eqnarray}
where in the second step  expression (\ref{38}) for $X$ was used.
Using Eqs. (\ref{53}) and (\ref{62}) we have
\begin{equation}\label{65}
    k=-(1+A-D)\sigma_y\,.
\end{equation}
where
$\zeta^\dag\sigma_z\zeta=D\sigma_y$ was used. Now the calculation of
the terms entering the right hand side of Eq. (\ref{64}) is
straightforward:
\begin{eqnarray}\label{66}
  \frac{1}{2}\Str\left(\hat{X}k\right)^2&=&(1+2A-2D)(\tau^2-v^2)\,,\nonumber\\
  \frac{1}{2}\Str\hat{X}^2&=& \tau^2+v^2\,.
\end{eqnarray}
Collecting the results of this subsection, we have
\begin{eqnarray}\label{67}
 && \frac{1}{8}\Str \left[({\bf1}+2X){\bf
    K_1}\right]^2=\nonumber\\
    &&\tau(2u+1-\tau)-v^2+2(A-D)(\tau^2-v^2)\,.
\end{eqnarray}
whence follows Eq. (\ref{68}).

\end{document}